\documentclass[letter]{aa}
\usepackage{graphicx}
\usepackage{txfonts}
\begin{document}
\title{The occultation events of the Herbig Ae/Be star V1247~Ori}  
\titlerunning{The occultation events of the Herbig Ae/Be star V1247~Ori}
%
%\subtitle{...}
%
\author{Jos\'e A. Caballero\inst{1,2}}
\institute{
Centro de Astrobiolog\'{\i}a (CSIC-INTA), Carretera de Ajalvir km~4,
28850 Torrej\'on de Ardoz, Madrid, Spain
\and
Departamento de Astrof\'{\i}sica y Ciencias de la Atm\'osfera, Facultad de
F\'{\i}sica, Universidad Complutense de Madrid, 28040 Madrid, Spain,
\email{caballero@astrax.fis.ucm.es}}
\date{Received 7 January 2010; accepted 16 February 2010}

% \abstract{}{}{}{}{} 
% 5 {} token are mandatory
\abstract
{} 
{I~study new deep ($\Delta V \approx$ 1.20--1.65\,mag) occultation events of the
$\delta$~Scuti, Herbig Ae/Be star V1247~Ori in the Ori~OB1\,b association.}  
{I~use the $V$-band ASAS light curve of V1247~Ori, which covers the last nine
years, together with photometric data in the near-ultraviolet, visible, near-,
and far-infrared taken from the literature.
I~carry out a periodogram analysis of the ``cleaned'' light curve and construct
the spectral energy distribution of the star.}     
{The star V1247~Ori is interesting for the study of the UX~Orionis phenomenon,
in which Herbig Ae/Be stars are occulted by their protoplanetary discs, for
three reasons: brightness ($V \approx$ 9.85\,mag), large infrared excess at
20--100\,$\mu$m ($F_{60} \approx$ 10\,Jy), and photometric stability out of
occultation ($\sigma(V) \approx$ 0.02\,mag), which may help to determine the
location and spatial structure of the occulting disc clumps.}  
{}
\maketitle
%
%________________________________________________________________

\section{Introduction}
\label{introduction}

\begin{figure}
\centering
\includegraphics[width=0.49\textwidth]{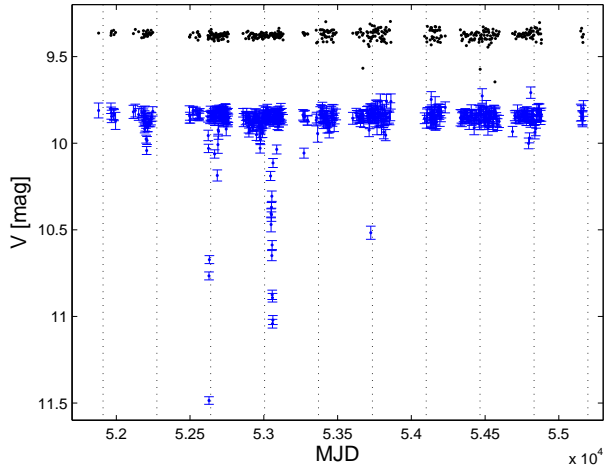}
\caption{ASAS light curves of V1247~Ori (bottom, [blue] error bars)
and a nearby, slightly brighter, field late-K star for comparison
(\object{HD~290760}, $\rho \sim$ 19\,arcmin; top, [black] dots).
Vertical dotted lines indicate the first day of the years 2001 to 2010.
The two main occultations events in V1247~Ori occurred in 2002~Dec
($V_{\rm min} \approx$ 11.50\,mag) and 2004~Feb ($V_{\rm min} \approx$
11.05\,mag). 
Note the seasonal gaps.}   
\label{figure.VvsMJD_errorbars}
% ./Software/sablelaser.m
\end{figure}

With the aim of finding highly variable stars of the Orion Belt not listed by
Caballero et~al. (2010), I~explored the All Sky Automated Survey ASAS-3
Photometric $V$-band Catalogue of variable stars of Pojma\'nski (2002). 
Among the light curves of variable ASAS stars unidentified by Caballero et~al.
(2010), the one of the pre-main sequence star \object{V1247~Ori} stood out
because of at least two unnoticed deep occultation events, one of which lasted
for about 20\,d (Fig.~\ref{figure.VvsMJD_errorbars}). 
Apart from that, the light curve of V1247~Ori is highly stable, with reported
light curve standard deviations of only 0.010--0.013\,mag (see below). 

The star V1247~Ori (HD~290764)\footnote{In some works, V1247~Ori is incorrectly
listed as BD--01~983 (e.g., Kazarotevs 1993; Rodr\'{\i}guez et~al. 1994), but
that is the B8 star \object{HD~290765}, and as vdB~50 (Vieira et~al. 2003), but
that is the B3Vn star \object{HD~37674}.} 
was classified as an intermediate A-type giant in
the Orion ``ring'' by Schild \& Cowley (1971). 
Later, Guetter (1981) assigned it membership in the Warren \& Hesser (1977)'s
subgroup b2, which is now known as the central part of the \object{Ori~OB1\,b}
association, in the vicinity of the bright supergiant \object{Alnilam}
($\epsilon$~Ori, $\tau \sim$ 5--10\,Ma -- Caballero \& Solano 2008). 
MacConnell (1982) discovered H$\alpha$ emission with an estimated
densitity 2 (``on a scale from T, trace, to 5'') in an optical spectrum of
V1247~Ori. 
Later, Vieira et al. (2003) identified an ``H$\alpha$ symmetric profile without,
or with only very shallow, absorption features'', but found no forbidden lines.  
Wouterloot \& Walmsley (1986) firstly identified the star with the infrared
source IRAS~05355--0117 and provided a restrictive upper limit to the presence
of H$_2$O radio lines.
Next Wouterloot et~al. (1988, 1989) did the same for NH$_3$ and CO radio lines.
The spectral energy distribution (SED) of V1247~Ori shows clear flux excess from
the $J$ band to 60--100\,$\mu$m (Fujii et~al. 2002; Clarke et~al. 2005;
Caballero \& Solano 2008), and is composed of two components, one warm
(1.2--2.2\,$\mu$m) and the other cool (20--100\,$\mu$m).  
It also displays intrinsic polarisation ($P_{\rm pol}$ =
0.32$\pm$0.04\,\%,  $\theta_{\rm pol}$ = 52.9\,deg), which indicates that its
circumstellar envelope is not spherical (Rodrigues et~al. 2009). 
Taking into account all these factors, V1247~Ori has been classified as a
Herbig Ae/Be star with a circumstellar disc (Herbig Ae/Be stars are the
high-mass ``counterparts'' of T~Tauri stars -- Wouterloot \& Walmsley 1986;
Garc\'{\i}a-Lario et~al. 1997; Fujii et~al. 2002; Vieira et~al. 2003). 

%______________________________________________ Figure 

Lampens \& Rufener (1982) repeatedly observed V1247~Ori on 189 occasions during
67 days in early 1984 and another 18 times on a night about one year later,
probably on 1984~Dec~03.
They indicated that the measurements made on that night were fainter by
0.03\,mag compared to the rest of the data. 
They found a period of photometric variability $P$ = 0.096967\,d with an
amplitude of only 0.016\,$\pm$\,0.008\,mag, and classified it as a
$\delta$~Scuti star. 
Afterwards, V1247~Ori has appeared in a number of catalogues of such kind of
pulsators (e.g., Garc\'{\i}a et~al. 1995; Rodr\'{\i}guez et~al. 2000). 
Furthermore, it is one of the few known pre-main-sequence
$\delta$~Scuti stars (Zwintz 2008). 
The actual period of photometric variability may slightly differ from the
Lampens \& Rufener (1982) value, since Debosscher et~al. (2007) tabulated
a different period at $P \approx$ 0.089973\,d.
Besides, the authors provided a second period at $P \approx$ 75.255\,d.
In any case, the measured amplitudes are very low, and the $V$ magnitude of
V1247~Ori has never been reported to vary more than 0.05\,mag. 
However, the occultation events in the ASAS light curve, which may be related to
an edge-on circumstellar disc, were deeper than 1\,mag.

\section{Analysis and results}
\label{analysisandresults}

%______________________________________________ Figure 
\begin{figure}
\centering
\includegraphics[width=0.49\textwidth]{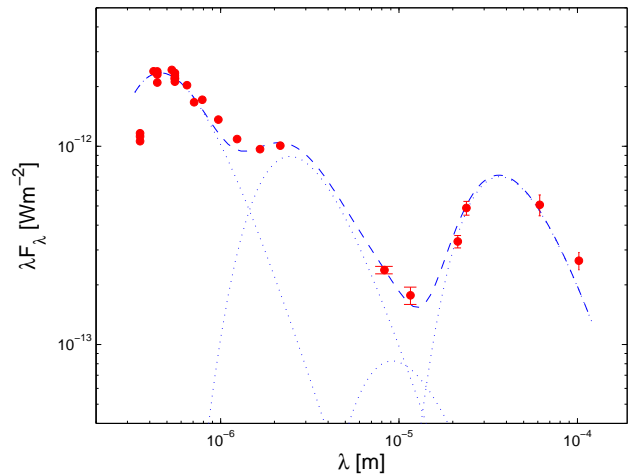}
\caption{Spectral energy distribution of V1247~Ori.
Data points correspond, from left to right, to $U$, $B_T$, $B$, $V_T$, $V$,
$R_C$, $R$, $I_C$, $I$, $J$, $H$, $K_{\rm s}$, and 8, 12, 21, 25, 60, and
100\,$\mu$m.
Only as a guidance and without any fitting purpose, four {\em spherical} black
bodies of $T_{\rm eff}$ = 8000, 1500, 400, and 100\,K are plotted with dotted
(blue) lines. 
The result of combining the four black bodies is marked with a dashed (blue)
line.}  
\label{figure.vf_lambdaFvslambda}
% ./Software/sed_V1247Ori.m 
\end{figure}

In Table~\ref{table.basicdata} I~compiled coordinates, proper
motions, magnitudes and fluxes (from the near-ultraviolet $U$ band to the {\em
IRAS} 100\,$\mu$m one), and spectral types of V1247~Ori from the literature.
Apart from the spectral type determinations by Schild \& Cowley (1971) and
Vieira et~al. (2003), there have been other measurements at A2 by Heckmann (1975
-- taken from the Publications of the Leander McCormick Observatory) and A7 by
Nesterov et~al. (1995).
I~tabulated neither the magnitudes $r'$ from the Carlsberg Meridian
Catalogue 14 (CMC14; Carlsberg Meridian Telescope 2006) and $i'$ from the Deep
Near Infrared Survey of the Southern Sky (DENIS; Epchtein et~al. 1997), which
lack suitable uncertainties, nor the magnitudes $B_J R_F I_N$ from digitised
photographic plates (e.g., USNO-B1; Monet et~al. 2003). 
Besides, I~have not taken into account the $JHK$ magnitudes provided by
Garc\'{\i}a-Lario et~al. (1997) and Doering \& Meixner (2009).
In Table~\ref{table.V} I~give eleven accurate measurements of the $V$
magnitude of V1247~Ori, as published in the literature. 
The mean and standard deviation of all the measurements are $\overline{V}$ =
9.861 and $\sigma(V)$ = 0.033\,mag.
With the information at the 18 wavelength bands in Table~\ref{table.basicdata},
I~constructed the SED shown in Fig.~\ref{figure.vf_lambdaFvslambda}. 
The infrared flux excess redwards of 1\,$\mu$m is obvious.

%__________________________________________________ One column table
   \begin{table}
      \caption[]{$V$ magnitudes of V1247~Ori published in the literature.} 
         \label{table.V}
     $$ 
         \begin{tabular}{cl c cl}
            \hline
            \hline
            \noalign{\smallskip}
$V$  			& Ref.$^{a}$ 	&	& $V$  			& Ref.$^{a}$ 	\\
{[mag]}  		& 		&	& {[mag]}  		& 		\\
            \noalign{\smallskip}
            \hline
            \noalign{\smallskip}
9.88 $\pm$ 0.01:	& Sh62		&	& 9.82 $\pm$ 0.03	& TYC   \\ 
9.854 $\pm$ 0.007	& Ru76		&	& 9.87 $\pm$ 0.02	& Ri00  \\
9.846 $\pm$ 0.013	& Ru88		&	& 9.81 $\pm$ 0.05	& Fu02  \\
9.844 $\pm$ 0.010	& LR92		&	& 9.88 $\pm$ 0.01:	& Vi03	\\
9.90 $\pm$ 0.01		& Eg92		&	& 9.85 $\pm$ 0.13	& Dr06	\\
9.92 $\pm$ 0.01: 	& Ha99		\\ % $\pm$ 0.006 
         \noalign{\smallskip}
            \hline
         \end{tabular}
     $$ 
\begin{list}{}{}
\item[$^{a}$] References --
Sh62: Sharpless 1962;
Ru76: Rufener 1976;
Ru88: Rufener 1988;
LR92: Lampens \& Rufener 1992;
Eg92: Egret et~al. 1992;
Ha99: Handler 1999;
TYC: H{\o}g et~al. 2000, transformed from $B_T V_T$ using the expression $V =
V_T - 0.090 (B_T - V_T)$;
Ri00: Richmond et~al. 2000;
Fu02: Fujii et~al. 2002;
Vi03: Vieira et~al. 2003;
Dr06: Droege et~al. 2006.
\end{list}
   \end{table}
%

%______________________________________________ Figure 
\begin{figure*}
\centering
\includegraphics[width=0.49\textwidth]{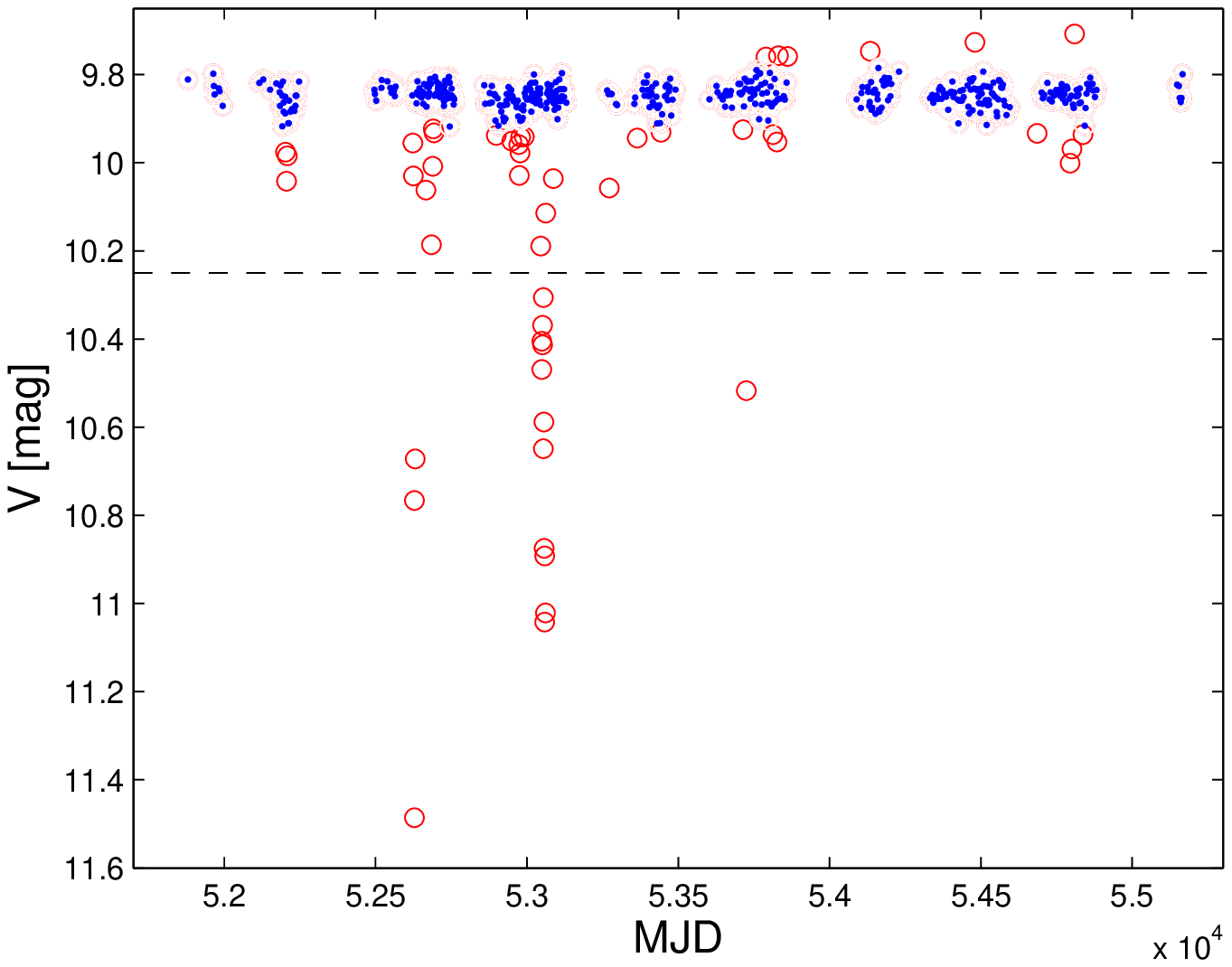}
\includegraphics[width=0.49\textwidth]{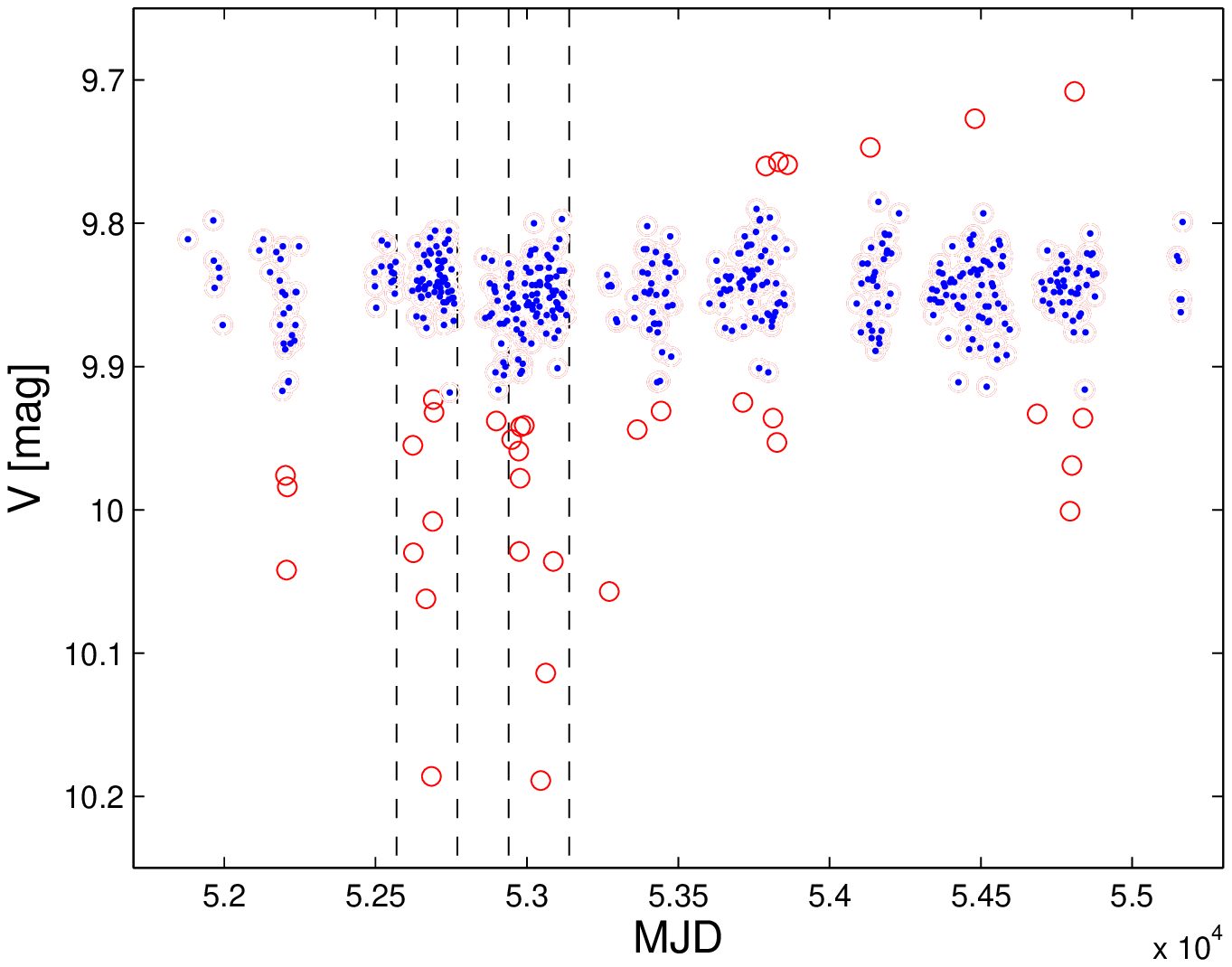}
\includegraphics[width=0.49\textwidth]{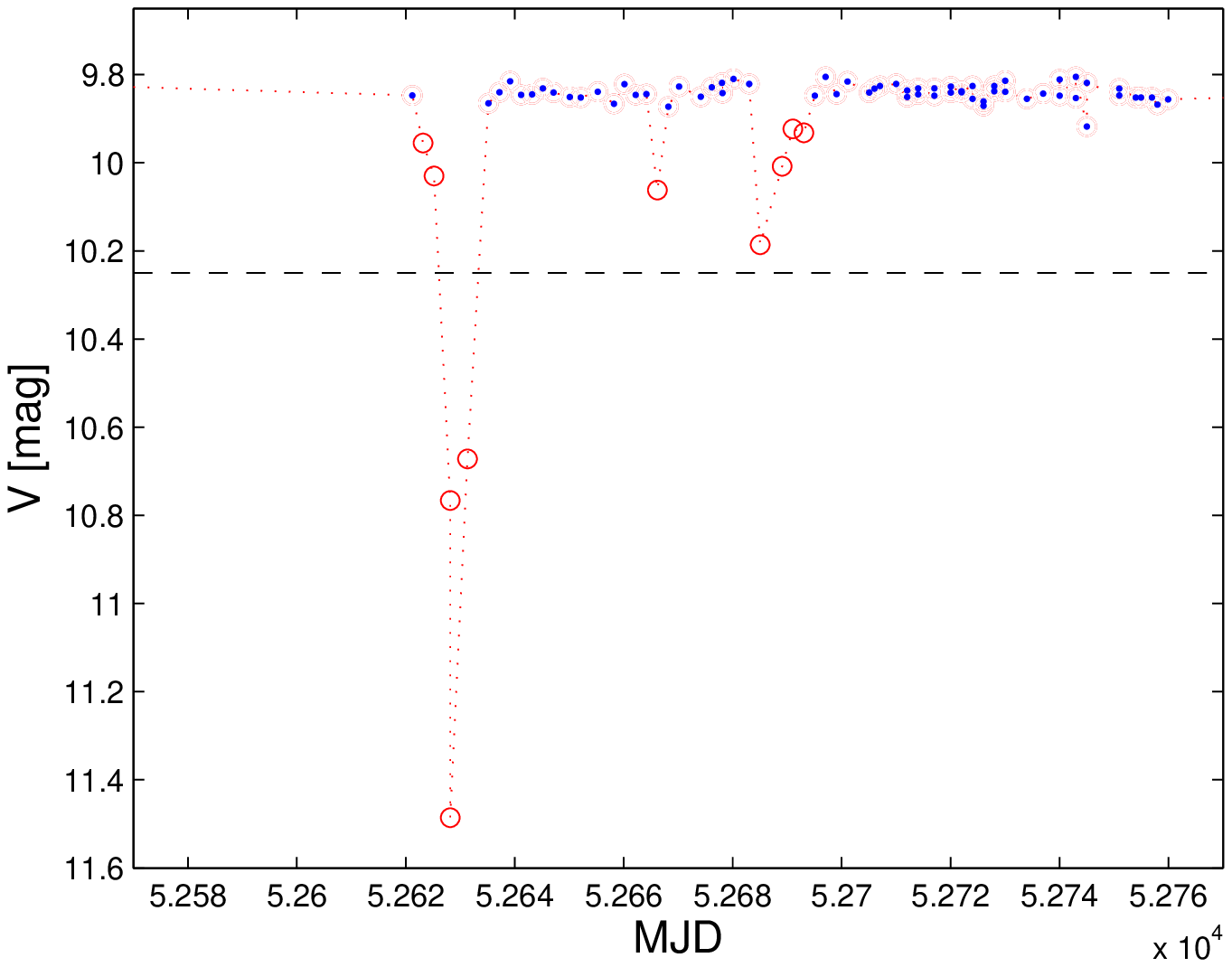}
\includegraphics[width=0.49\textwidth]{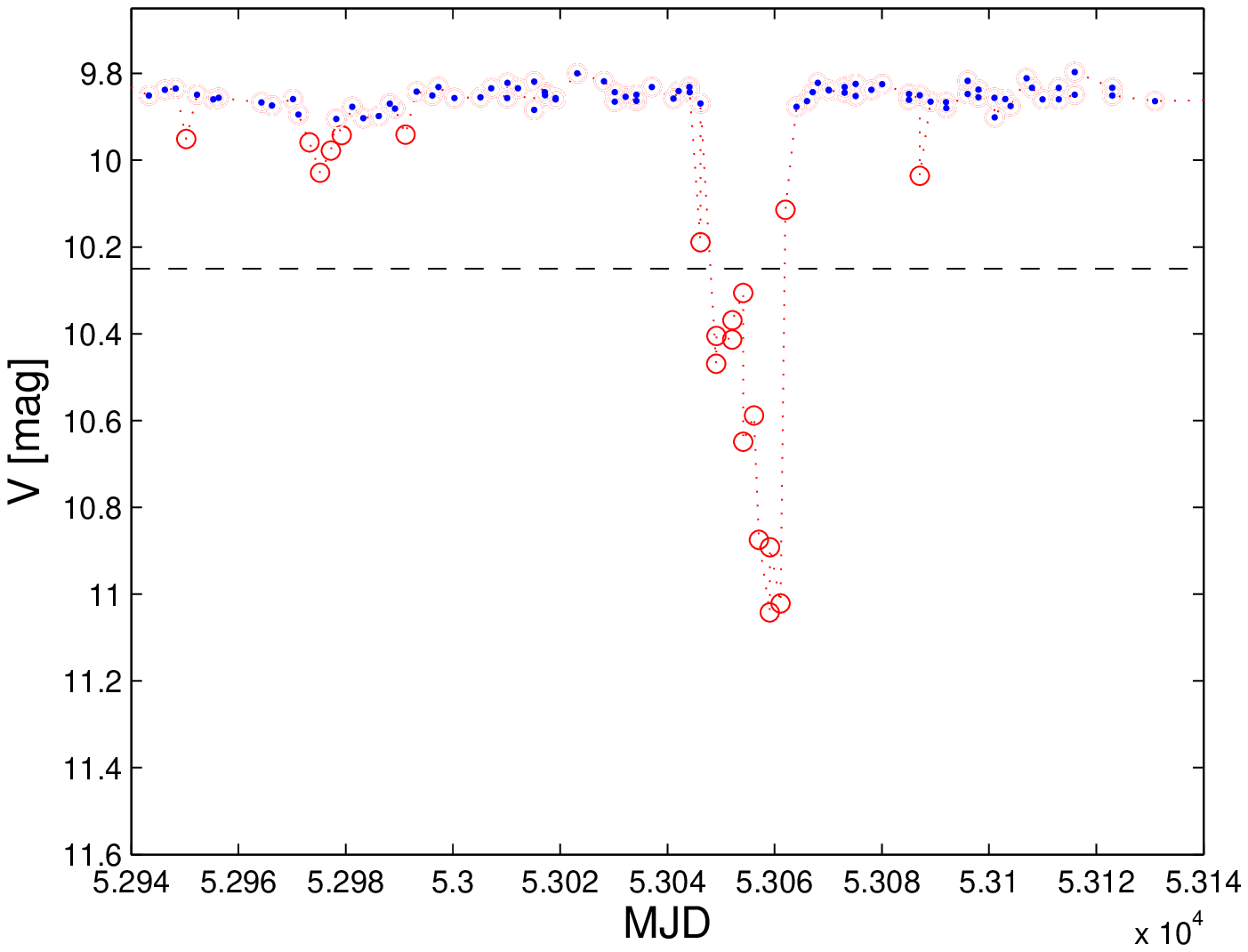}
\caption{Different captions of the ASAS light curve of V1247~Ori.
Data points that deviate less and more than $3\sigma$ from the
iteratively-``cleaned'' light curve are plotted with small (blue) dots and open
(red) circles, respectively.
{\em Top left panel:} full time and magnitude ranges. 
The horizontal dashed line marks the value $V$ = 10.25\,mag.
{\em Top right panel:} full time range.
Only magnitudes brighter than $V$ = 10.25\,mag are shown.
The vertical dashed lines indicate the enlargement areas in the bottom panels,
corresponding to the two main occultation events. 
{\em Bottom panels:} full magnitude ranges, 200\,d-wide time intervals around
the 2002~Dec ({\em left}) and 2004~Feb ({\em right}) occultation events.}   
\label{figure.VvsMJD_a-d}
% ./Software/sablelaser.m
\end{figure*}

The light curves in Fig.~\ref{figure.VvsMJD_errorbars} were built using
only the ASAS data points corresponding to the aperture ``2'' (i.e., 4 pixels, or
about 1\,arcmin -- Pojma\'nski 2002; Caballero et~al. 2010), which gave the
lower mean photometric error.
Besides, only data points with the best quality (grades ``A'' and ``B'') were
used. 
The ASAS photometry of V1247~Ori was not affected by neighbouring bright stars
(Fig.~\ref{figure.vf_ds9_bw}), and thus the observed photometric variability is
intrinsic to the Herbig Ae/Be star. 
For a better visualisation of the occultation events, some captions of the light
curve of V1247~Ori are shown in the four panels of
Fig.~\ref{figure.VvsMJD_a-d}. 
I~implemented an iterative process to differentiate the data points that
deviate more than $3\sigma$ from a ``cleaned'' light curve with the rest of the
data points.
The mean and standard deviation of the ``cleaned'' light curve, indicated with
small dots in Fig.~\ref{figure.VvsMJD_a-d}, are $\overline{V}$ = 9.847\,mag and
$\sigma(V)$ = 0.024\,mag, consistent with the values given in 
Table~\ref{table.V}.

\section{Discussion}
\label{discussion}

The ASAS light curve of V1247~Ori has 509 data points and covers about nine
years with short seasonal gaps.
Only 51 data points, marked with open circles in Fig.~\ref{figure.VvsMJD_a-d},
deviate more than $3\sigma$ with respect to the ``cleaned'' light curve.
I~carried out a periodogram analysis on the other 458 data points of the
``cleaned'' light curve.
The analysis, identical to the one in Caballero et~al. (2010), gave a power
maximum at 560\,d, which is an alias of the total time span (one sixth).
The second most powerful peak is much narrower and falls at $P \approx$ 76.5\,d.
It likely corresponds to the $P_2 \approx$ 75.255\,d periodicity found by
Debosscher et~al. (2007).
Given the typical time interval between consecutive ASAS observations (about
2--3\,d; see again Caballero et~al. 2010), this analysis is not sensitive to the
short $\delta$~Scuti periodicities of 2.16--2.33\,h found by Lampens \&
Rufener (1990) and Debosscher et~al. (2007).

There are two kinds of outlier data points (circles in
Fig.~\ref{figure.VvsMJD_a-d}). 
Six of the 51 outliers are brighter than the ``cleaned'' light curve and are
likely associated to flare-like events.
There were three such events in the season 2005--2006, but only one in each
of the following three seasons.
The light curve hints at an increasing amplitude of the flare-like events with
time, from about 0.10\,mag in 2006 to about 0.15\,mag in 2009.
They follow one after other with about the same time interval of about one
year (note that each ``flare'' is a single event that was observed only once
per night).
The light curve of the stable field star HD~290760 in
Fig.~\ref{figure.VvsMJD_errorbars} has the same standard deviation as the
``cleaned'' light curve of V1247~Ori, and shows three outlier data points (of a
total of 459).
From it, I~estimate that only about 0.7\,\% of the data points of the light
curve of V1247~Ori can be affected by systematics of unknown origin and that
thus most of the flare-like events can be real. 
They can be related to the H$\alpha$ emission discovered by McConnell (1982) and
probably to on-going accretion from the circumstellar disc of V1247~Ori.
The other 45 outlier data points are fainter than the ``cleaned'' light curve.
Most of them were associated to the main occultation events in 2002~Dec and
2004~Feb, but there were also secondary events of less depth containing groups
of up to four data points (e.g., 2001~Oct [+0.20\,mag], 2003~Feb [+0.20\,mag],
2003~Dec [+0.35\,mag], and 2008~Nov-Dec [+0.15\,mag]).
Besides, there was an isolated data point in 2005~Dec that was about 0.65\,mag
fainter than the mean ``cleaned'' light curve.

The two main occultation events are different in shape (bottom panels in
Fig.~\ref{figure.VvsMJD_a-d}).  
On the one hand, the 2002~Dec event lasted for about two weeks and was quite
symmetric, with a pronounced drop and raise around the minimum on 2002~Dec~20.
Because of the different slopes of the drop and raise, the event could have been
deeper than 1.65\,mag on 2002~Dec~21--23. 
On the other hand, the 2004~Feb event was not so deep (+1.20\,mag), lasted
longer (about three weeks), and was highly asymmetric, with a series of local
magnitude maxima and minima. 
While the origin of the first event might have been a transiting stellar
companion, the one of the 2004~Feb event must be associated to an heterogeneous
envelope, which is likely the circumstellar disc (a sudden burst of cool spots
in the photosphere of an early-type star would be rather speculative).

From the SED in Fig.~\ref{figure.vf_lambdaFvslambda}, I~confirm that the disc of
V1247~Ori has at least two components at different temperatures.
The simplest model indicates that there is a warm ($T \sim$ 1500\,K) inner
component and a cool ($T \sim$ 100\,K) outer one, with a dim third component at
an intermediate temperature ($T \sim$ 400\,K) in between. 
Obviously, protoplanetary discs are not isothermal spheres, and the black bodies
in Fig.~\ref{figure.vf_lambdaFvslambda} must be understood only as a guidance. 
A more comprehensive analysis, accounting for the disc inner rim, shadow region,
and the disc atmosphere and surface beyond the flaring radius should be used to
derive accurate disc properties (e.g., radius and height of the rim, surface
density, total disc size, disc luminosity -- Dullemond et~al. 2001). 
The high flux at 20--60\,$\mu$m reveals a large physical size of the outer-disc
component.

Although occultations by circumstellar discs of young stellar objects in
general and of Herbig Ae/Be stars in particular are not rare (e.g., Bibo \&
Th\'e 1991; Grinin et~al. 1991; Herbst et~al. 1994; Bertout 2000; Dullemond
et~al. 2003), the case of V1247~Ori sheds bright light on the topic for
several reasons:
\begin{itemize}

\item The variability type of Herbig Ae/Be stars with occulting edge-on discs is
called UX~Orionis after the best-observed star displaying this sort of activity.
V1247~Ori is as bright as \object{UX~Ori} itself ($V \approx$ 9.6\,mag), and
more than 4\,mag brighter than the interesting eclipsed star \object{KH~15D}
(Hamilton et~al. 2005).

\item The SED of V1247~Ori is well characterised.
The two models of protoplanetary discs that best explain the UX~Orionis
phenomenon are based on the occultation by clumps of dust and gas that cross
our line of sight.
These hydrodynamics fluctuations may be located in the outer-disc regions or in
the puffed-up inner rim (Dullemond et~al. 2003).
The second location favours self-shadowed discs with relatively weak
far-infrared excess.
The red {\em IRAS} colour $[12]-[60] \approx$ 3.0\,mag and strong far-infrared
excess of V1247~Ori supports instead the first location.

\item Most interestingly, the stability of the bright phase of V1247~Ori, at the
level of 0.02\,mag or less, indicates that the variability observed in the
2004~Feb event must be exclusively ascribed to the occulting disc clumps and not
to photospheric variations.
Thus, the spatial structure of the hydrodynamic fluctuations in the inner rim
or the outer disc (maybe spiral arm-like) can be derived from the shape of the
light curve with an appropriate analysis.

\end{itemize}

However, there might be further complications, such as stellar
multiplicity. The SED of V1247~Ori has a gap around about 15\,$\mu$m,
which is frequently considered as a signature of binarity in a young star
(e.g., Espaillat et~al. 2007).
If the star is a binary, the structure of the circumstellar disc around the
primary can differ from the puffed-up inner rim model. 
This complication would make it difficult to localise the dust clouds screening
the star from the observer.
In addition, following the ideas shown by Grinin \& Rostopchina (1996), the
unusual photometric properties of V1247~Ori can be caused by the intermediate
orientation of its circumstellar disc, between pole-on (corresponding to active
UX~Ori stars with two-peaked H$\alpha$ profiles) and edge-on (corresponding to
non-active Herbig Ae/Be stars with P~Cygni H$\alpha$ profiles and Algol-type
activity like \object{AB~Aur}).
Further photometric monitoring of V1247~Ori is required to ascertain the
structure of its disc.

\begin{acknowledgements}

I thank V.~P.~Grinin for his helpful referee report.
I~am an {\em investigador Ram\'on y Cajal} at the Centro de Astrobiolog\'{\i}a. 
This research made use of the SIMBAD, operated at Centre de Donn\'ees
astronomiques de Strasbourg, France, and NASA's Astrophysics Data System.
Financial support was provided by the Universidad Complutense de Madrid and the
Spanish Ministerio de Ciencia e Innovaci\'on under grants
AyA2008-06423-C03-03 and 		% (WSO) 
AyA2008-00695.				% (estrellax II)

\end{acknowledgements}

\appendix

\section{Basic data and finding chart of the Herbig Ae/Be star V1247~Ori}

%__________________________________________________ One column table
   \begin{table}
      \caption[]{Basic data of the Herbig Ae/Be star V1247~Ori.} 
         \label{table.basicdata}
     $$ 
         \begin{tabular}{l c l l}
            \hline
            \hline
            \noalign{\smallskip}
Quantity  			& Value			& Unit		& Reference$^{a}$ \\
            \noalign{\smallskip}
            \hline
            \noalign{\smallskip}
$\alpha^{J2000}$		& 05 38 05.25		&		& TYC	\\
$\delta^{J2000}$		& --01 15 21.7		&		& TYC	\\
$\mu_\alpha \cos{\delta}$	& --0.9 $\pm$ 1.1	& mas\,a$^{-1}$	& TYC	\\
$\mu_\delta$			&  +1.3 $\pm$ 1.1	& mas\,a$^{-1}$	& TYC	\\
$U$				& 10.29 $\pm$ 0.01:	& mag		& Sh62	\\
$B$				& 10.20 $\pm$ 0.01:	& mag		& Sh62	\\
$B_T$				& 10.18 $\pm$ 0.03	& mag		& TYC	\\
$V$				&  9.88 $\pm$ 0.01:	& mag		& Sh62	\\
$V_T$				&  9.85 $\pm$ 0.03	& mag		& TYC	\\
%$r'$				&  9.771 		& mag		& CMC14	\\
$R$				&  9.68 $\pm$ 0.01:	& mag		& Vi03	\\
$R_C$				&  9.61 $\pm$ 0.05	& mag		& Fu02	\\
%$i'$				&  9.637 $\pm$ 1.0:	& mag		& DENIS	\\
$I$				&  9.44 $\pm$ 0.01:	& mag		& Vi03	\\
$I_C$				&  9.32 $\pm$ 0.05	& mag		& Fu02	\\
$J$				&  8.88 $\pm$ 0.03	& mag		& 2MASS	\\
$H$				&  8.20 $\pm$ 0.05	& mag		& 2MASS	\\
$K_{\rm s}$			&  7.41 $\pm$ 0.03	& mag		& 2MASS	\\
8\,$\mu$m			&  0.66 $\pm$ 0.03	& Jy		& MSX6C	\\
12\,$\mu$m			&  0.68 $\pm$ 0.07	& Jy		& IPAC	\\
21\,$\mu$m			&  2.36 $\pm$ 0.17	& Jy		& MSX6C	\\
25\,$\mu$m			&   3.9 $\pm$ 0.3	& Jy		& IPAC	\\
60\,$\mu$m			&  10.4 $\pm$ 1.2	& Jy		& IPAC	\\
100\,$\mu$m			&   9.0 $\pm$ 0.9	& Jy		& IPAC	\\
Sp. type			& A5~III, F0~Ve		& 		& SC71, Vi03	\\
         \noalign{\smallskip}
            \hline
         \end{tabular}
     $$ 
\begin{list}{}{}
\item[$^{a}$] References --
TYC: H{\o}g et~al. 2000;
Sh62: Sharpless 1962;
Vi03: Vieira et~al. 2003;
Fu02: Fujii et~al. 2002;
2MASS: Skrutskie et~al. 2006;
MSX6C: Egan et~al. 2003 (8 and 21\,$\mu$m-band measurements correspond to the
SPIRIT~III/{\em Midcourse Space Experiment} bands $A$, 8.28\,$\mu$m
[6.8--10.8\,$\mu$m], and $E$, 21.34\,$\mu$m [18.2--25.1\,$\mu$m];
IPAC: Joint {\em IRAS} Science Working Group 1988;
SC71: Schild \& Cowley 1971.
\end{list}
   \end{table}
%

%______________________________________________ Figure 
\begin{figure}
\centering
\includegraphics[width=0.49\textwidth]{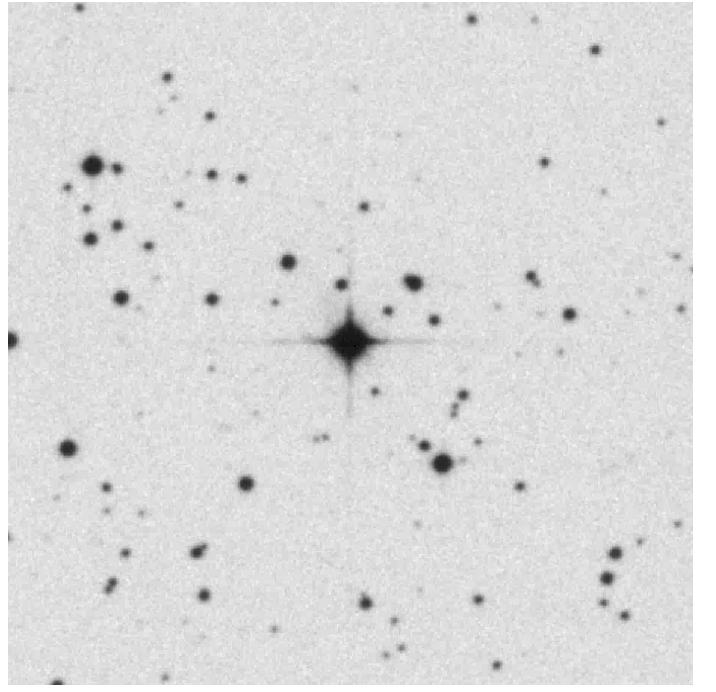}
\caption{Inverted-colour image centred on V1247~Ori from the SuperCOSMOS
digitisations of the United Kingdom Schmidt Telescope photographic red plates.
The size is 5\,arcmin\,$\times$\,5\,arcmin.
North is up and east is to the left.
The second and brightest stars in the field have magnitudes $V$ =
14.3$\pm$0.2 (at 1.1\,arcmin to the southwest of V1247~Ori) and
13.8$\pm$0.4\,mag (at 2.3\,arcmin to the northeast).}   
\label{figure.vf_ds9_bw}
% ./Software ... ds9
\end{figure}

\end{document}